\documentclass[12pt]{article}
\usepackage{epsfig}
\usepackage{rotating}
\usepackage{color}
\usepackage{longtable}
\usepackage{multicol}
\usepackage[normalem]{ulem}
\usepackage{hyperref}
\usepackage{breakurl}
\usepackage{graphicx}
\usepackage{authblk}
\usepackage{geometry}
\title{Human mobility patterns in Mexico City and their links with socioeconomic variables during the COVID-19 pandemic} 

\author{Oscar Fontanelli$^{1, \ast}$, Dulce I. Valdivia$^1$, Guillermo Romero$^2$, Oliver Medina$^3$, Wentian Li$^4$, Maribel Hernández-Rosales$^1$}
\date{
\small{
$^1$ Genetic Engineering Department, Centro de Investigación y de Estudios Avanzados del Instituto Politécnico Nacional, Unidad Irapuato, Mexico.\\
$^2$ Data-Pop Alliance, USA.\\
$^3$ Facultad de Ciencias Políticas y Sociales, Universidad Nacional Autónoma de México.\\
$^4$ The Robert S. Boas Center for Genomics and Human Genetics,
The Feinstein Institutes for Medical Research, NY. US.\\
$^\ast$ oscar.fontanelli@cinvestav.mx\\
\today
}
}

\begin{document}

\maketitle

Keywords: human mobility; mobility patterns; origin-destination networks; Mexico City;
COVID-19; degree distribution.

\begin{abstract}

The availability of cellphone geolocation data provides a remarkable opportunity to study human mobility patterns and how these patterns are affected by the recent pandemic. Two simple centrality metrics allow us to measure two different aspects of mobility in origin-destination networks constructed with this type of data: variety of places connected to a certain node (degree) and number of people that travel to or from a given node (strength). In this contribution, we present an analysis of node degree and strength in daily origin-destination networks for Greater Mexico City during 2020. Unlike what is observed in many complex networks, these origin-destination networks are not scale free. Instead, there is a characteristic scale defined by the distribution peak; centrality distributions exhibit a skewed two-tail distribution with power law decay on each side of the peak. We found that high mobility areas tend to be closer to the city center, have higher population and better socioeconomic conditions. Areas with anomalous behavior are almost always on the periphery of the city, where we can also observe qualitative difference in mobility patterns between east and west. Finally, we study the effect of mobility restrictions due to the outbreak of the COVID-19 pandemics on these mobility patterns. 

\end{abstract}

\section{Introduction}

Mexico City is one of the largest cities in the world. With approximately 21 million inhabitants, it is the largest metropolitan area in North America and among the five largest metropolitan areas around the globe. It is also one of the most important economic, political and cultural hubs in the Americas. Human mobility within a city like this is a very complex phenomenon and the identification and analysis of human mobility patterns is a topic of wide interest for policy makers, urban planning, transportation planning, epidemic control, among others \cite{barbosa2018human, caschili2013accessibility, de2010modeling, gonzalez2008understanding, 
lenormand2012universal, simini2012universal}.

One common and useful approach for modeling human mobility in a city or region is the construction and characterization of origin-destination networks. These are weighted networks where nodes represent geographical areas or points of interest in the city and edges show the number of travels from one area to another. These networks can be built from different sources, the most common ones are census data or mobility surveys \cite{caschili2013accessibility, de2010modeling,goetz2010us,ramasco2009using}; however, other data sources such as taxi trips data \cite{riascos2020networks}, GPS taxi data \cite{zheng2016two}, smart card fare data \cite{alsger2015use}, shared bike data \cite{lotero2016rich, loaiza2019human} or social network data \cite{osorio2019social,pourebrahim2019trip} have also been used for this purpose. More recently, various studies have utilized cellphone location data to estimate origin-destination matrices and build mobility networks \cite{akin2002estimating, becker2013human, bengtsson2015using, caceres2007deriving, calabrese2011estimating, ccolak2015analyzing, cavalli2020mobility,edsberg2022understanding,
fekih2021data,louail2015uncovering,jiang2017activity,
panigutti2017assessing}.

Analysis of the distribution and scaling laws of centrality measures of these complex networks has proven to be a useful approach to identify mobility patterns \cite{chowell2003scaling, halepovic2005characterizing, han2011origin, zhao2015explaining, zhao2008empirical, zheng2021extreme, zhou2013human}, and how these patterns have changed with the outbreak of COVID-19 has also been a topic of wide interest in the last months and years \cite{alessandretti2022human,benita2021human, bonaccorsi2020economic, fajgelbaum2021optimal, hu2021human, iacus2020human,kraemer2020effect, schlosser2020covid, seto2020commuting, smolyak2021effects, sun2020quantifying}. 

Furthermore, several studies have focused on the influence of geographic factors on human mobility \cite{kraemer2020mapping,luo2016explore, ruktanonchai2021practical, toole2015coupling}. For example, in a study in Rio de Janeiro, Brazil \cite{lima2019urban} the authors related the geography of the city to mobility metrics quantified by node centrality measures. On the other hand, it is interesting to study how mobility metrics are linked to social and economic variables; a related study, for the cases of Medellín and Manizales in Colombia is presented in \cite{lotero2016rich}. More examples on the relationships between human mobility and socioeconomic factors can be seen in \cite{barbosa2021uncovering, deng2021high, li2019uncovering, xu2018human}. 

Regarding Mexico City, a recent study on mobility patterns based on Points of Interest data from Google Places and a radiation model for human mobility is presented in \cite{melikov2021characterizing}, and another study on the unequal access and appropriation of urban space in Mexico City through mobility data is presented in \cite{Letouze2022Parallel}. 

In this work, we investigate mobility patterns and how they changed with the beginning of the COVID-19 pandemics in Mexico and the subsequent lockdowns and restrictions by analyzing daily origin-destination networks for Greater Mexico City during 2020 built from a large dataset of anonymized cellphone geolocation data using a methodology introduced in a previous work \cite{arXiv}. 

What can we learn about mobility patterns in Mexico City and their evolution during the first months of the pandemic by looking at node degree and node strength in these origin-destination networks? These two node centrality measures quantify two different aspects of mobility: node degree is a measure of the number of different places from where people arrive or to where people go from a certain region, while node strength measures the total number of people that travel to or from a given region. 

The distributions of degree and strength have a very definite shape that is neither a power law nor lognormal. This indicates that these origin-destination networks are not scale free and that there is a characteristic scale that can be defined through the distribution mode. We utilize the Beta Rank Function (BRF) distribution to model these distributions and show that it performs better than power-law and lognormal. BRF is a probability distribution with a smooth peak that decays as a different power law on each side. Furthermore, with this distribution we can see that our origin-destination networks have two distinct regimes or scales of behavior, divided by the peak, which we call ``high'' and ``low'' mobility nodes. Each city region then belongs to either the high or low regime.

With these origin-destination networks and their degree and strength distributions we address the following questions: what are the main mobility hubs in the network? Do they change with the beginning of the pandemic and subsequent containment measures? Are both metrics, degree and strength, always correlated? What social, economic or geographical variables explain the formation of the two mobility regimes? Which nodes switch from high to low mobility, or the other way around, with the pandemics? Are both mobility groups equally affected by the pandemics? Finally, did the city go back to pre-pandemic mobility patterns during the study period? 

\section{Preliminaries and background}

\subsection{Human origin-destination networks}

The National Institute for Statistics, Geography and Informatics in Mexico (INEGI) defines the Mexico City Metropolitan Area (Greater Mexico City) as an urban area encompassing Mexico City itself and 60 adjacent municipalities from two neighboring states, State of Mexico and Hidalgo. This same institution provides official information on administrative divisions in the country. The smallest such division is the AGEB (basic geo-statistic area). According to the official definition, an urban AGEB is a geographical area of perfectly delimited blocks with habitational, commercial or industrial land use and with a minimum population of 2500 inhabitants. Greater Mexico City has a total of 5705 urban AGEBs. 

In \cite{arXiv} we presented a methodology to construct mobility networks from a large and anonymized database of mobile phone location data and presented a few examples for Greater Mexico City. Using this methodology we obtained 366 daily networks for 2020, each one being weighted and directed, where nodes are urban AGEBs in the city, edges represent observed mobility from one node to another node; and weight on edges shows the aggregated number of mobile phones that traveled from node $i$ to node $j$. Due to variability in the raw data, network size shows slight variations from day to day. The average number of nodes is 5599, standard deviation of 16.8 and a coefficient of variation of 0.003; the average number of edges is 281709, standard deviation of 84,513 and a coefficient of variation of 0.3. This set of 366 networks is the subject of study in this work. 

Next we briefly describe some useful background for our degree and strength distribution analysis. 

\subsection{Beta Rank Function}

\begin{figure}[tp]
    \centering
    \includegraphics[width=0.6\linewidth]{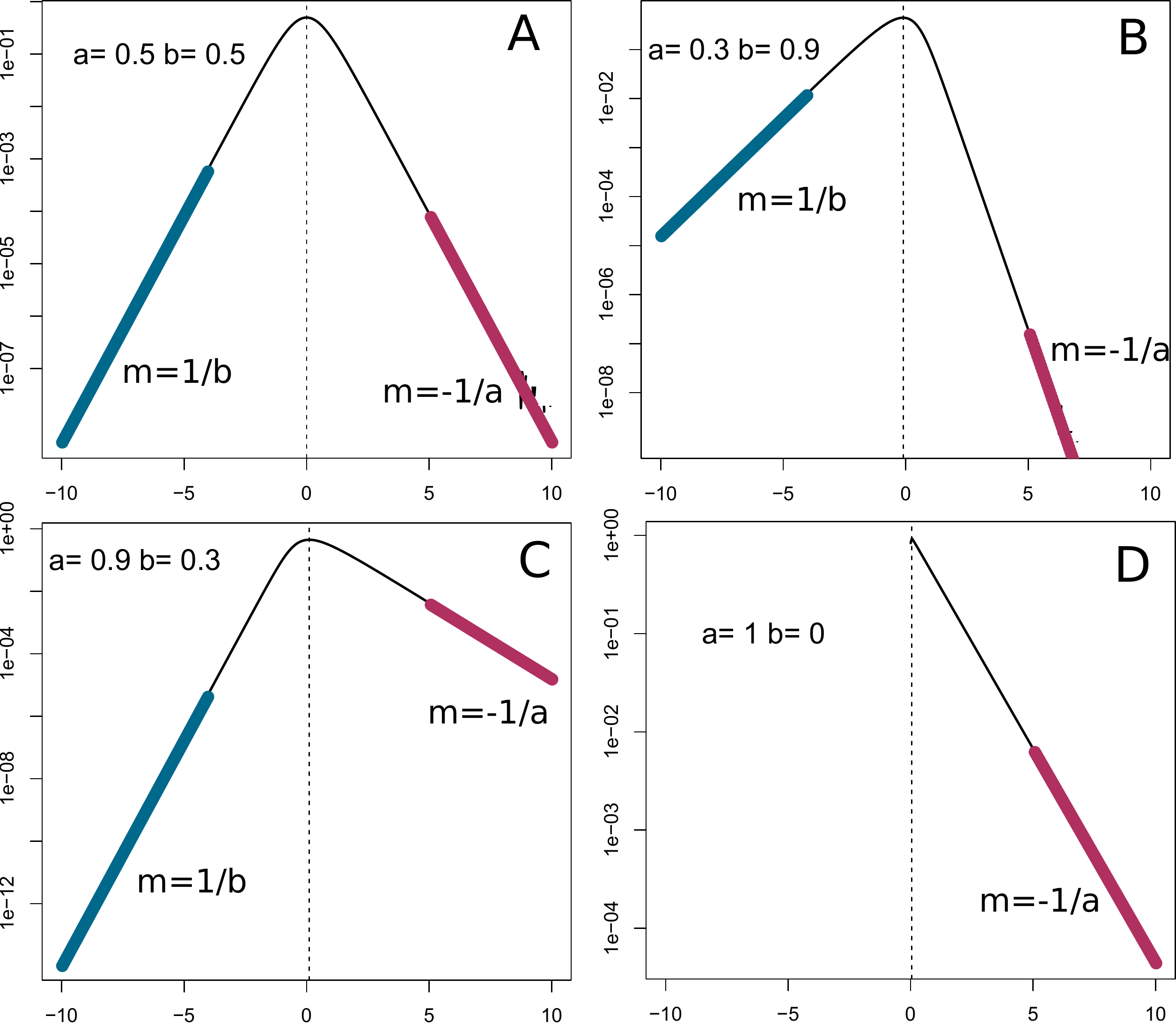}
    \caption{Semilog representation of the pdf of the logarithm of a BRF random variable. Decay on each side of the mode is controlled by an independent parameter, so the pdf can be asymmetric. Notice how the BRF reduces to power law when b=0.}
    \label{densities}
\end{figure}

A rank-size function is a function that gives the size of an observation $x$ in terms of its rank $r$, that is, the place it occupies on a decreasing ranked list. For example, the power law is the rank-size function $x(r) \sim 1/x^{\alpha}$, where $\alpha$ is the scaling parameter.

The Discrete Generalized Beta Distribution (DGBD) is the rank-size function given by

\begin{equation}
\label{DGBD}
x(r) = A\frac{(N+1-r)^b}{r^a},
\end{equation}

\noindent where $b$ and $a$ are free parameters, $N$ the maximum rank and $A$ a normalization constant \cite{martinez2009universality}. 

In \cite{fontanelli2020beta} we introduced the Beta Rank Function distribution (BRF), which is the probability distribution of a random variable whose rank-size function is the DGBD. This family of probability distributions can be defined through its quantile function $x(u)=A(1-u)^b/u^a$ \cite{hankin2006new}. When a random variable $X$ with a BRF distribution is log-transformed, the resulting random variable $Z=\log X$ has a probability density function (pdf) with a smooth peak and with an exponential decay on both tails.


This behavior can be geometrically observed if we plot the pdf of Z in semilog-representation. 
See in Fig.\ref{densities} how the pdf has a smooth peak and it decays as a straight line (in semilog-representation), where parameters $b$ and $a$ control the slope of the decay on the left and right tail respectively (slope on the left and right sides are $1/b$ and $-1/a$ respectively). If $a=b$, the pdf is symmetric (Fig. \ref{densities}-A); if $a>b$ the right tail is heavier than the left tail and the other way around (Fig. \ref{densities}-B and C). Notice how the BRF reduces to the power law when $b=0$, which can be seen in Eq. \ref{DGBD} and in Fig.\ref{densities}-D. Finally, the BRF distribution has finite moments of order $n<1/a$, which means that parameter $a$ controls the weight of the distribution tail, just as in a power law or a Pareto distribution. 

This BRF distribution is essentially a double power law (or double Pareto) with a smooth peak, heavy tails and power law decays on each side of the peak (possibly with different exponents on each side). 

In this work, we propose the BRF distribution to fit and model degree and strength distributions for the set of origin-destination networks. 

\section{Results}

\subsection{In- and out-degree, in- and out-strength are highly correlated}

Recall that origin-destination networks are weighted and directed networks, where weight of link $(i,j)$ indicates the number of observed travels from node $i$ to node $j$, in that direction. In directed networks, one must distinguish between in- and out-degree, and in- and out-strength, arising a natural question: 
are these in and out metrics correlated in these networks? For each network we computed the linear correlation coefficients between in-degree and out-degree, and between in-strength and out-strength. For all 366 days, we found all correlations coefficients above 0.95 for degree, and above 0.99 for strength. Additionally, we fitted linear models \emph{in-degree} $\sim$ \emph{out-degree} and \emph{in-strength} $\sim$ \emph{out-strength}. In all cases we found determination coefficients ($R^2$) above 0.99. For the vast majority of cases, in and out metrics are essentially the same, except for anomalous nodes. 

\begin{figure}[!t]
    \centering
    \includegraphics[width=1\linewidth]{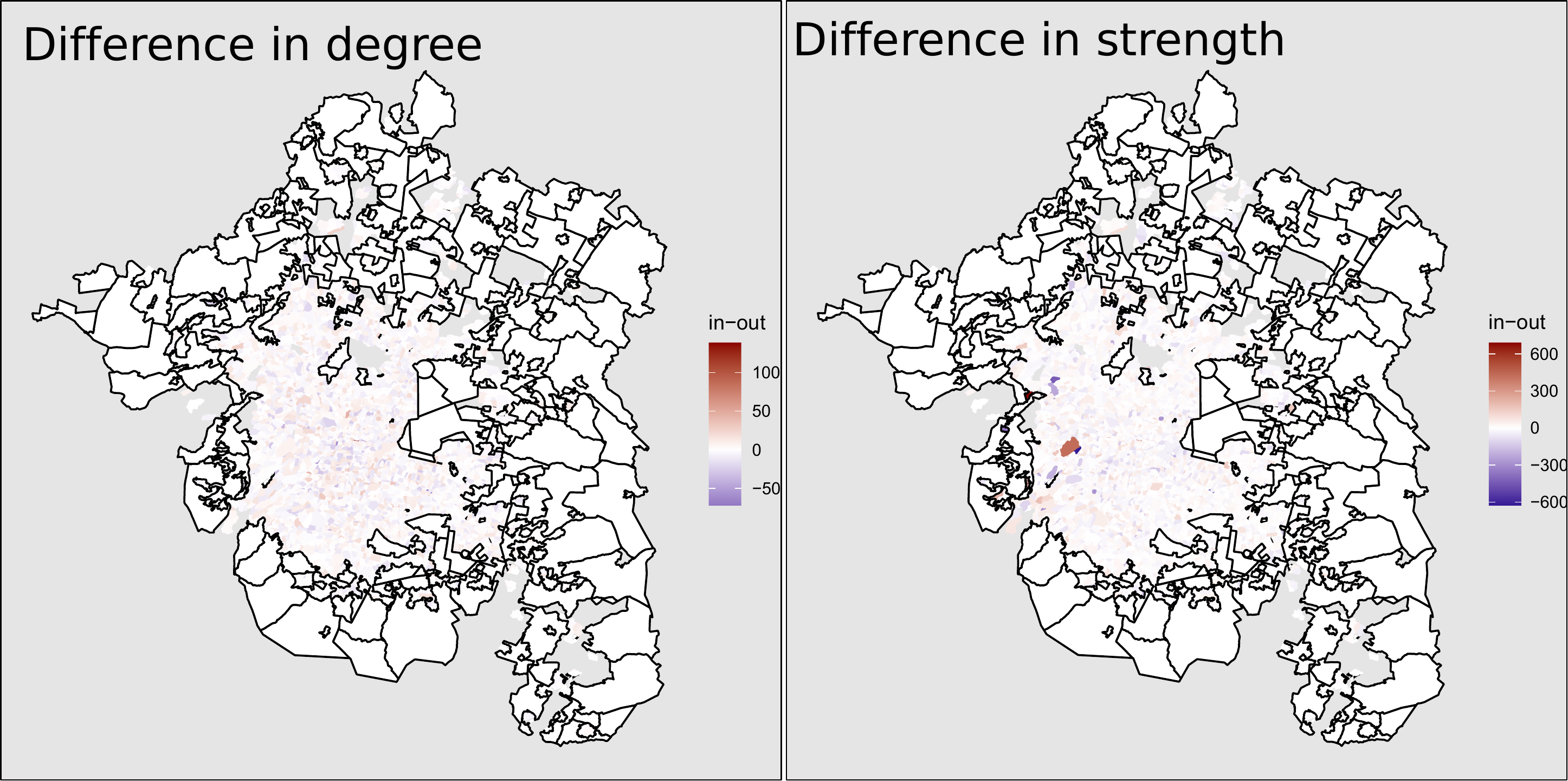}
    \caption{AGEBs in the city (network nodes) colored according to difference (in-degree)-(out-degree) and (in-strength)-(out-strength) for the February 17th network. }
    \label{fig::in_out}
\end{figure}

In general, we do not observe that exceptional nodes with high variations between in- and out-degree are geographically close. On the other hand, atypical nodes with high differences between in- and out-strength tend to be together and to occur on the periphery of the city. We show an example in Fig. \ref{fig::in_out}, which corresponds to the February 17 network. As we can see, there is a blue zone (out-strength $>$ in-strength) and a red zone (in-strength $>$ out-strength), both on the west side of the city. 

From now on, we analyze \emph{total} degree and strength (in $+$ out) and we refer to them simply as \emph{degree} and \emph{strength}.

\subsection{There are two evolving regimes of human mobility behavior}

In order to find the function that best fits in- and out- degree and in- and out-strenght distributions, 
we compared lognormal, power law, and BRF by means of Kolmogorov Smirnov (KS) tests, and Akaike Information Criterion (AIC). Lognormal distribution is discarded in all cases due to significant deviation at both tails. For degree distribution, BRF has a lower KS statistic and a better AIC than power law in all 366 networks. For strength distributions, BRF has a better (lower) AIC than power law for all 366 days, but power law has an equal or better (equal or lower) KS statistic than BRF for nine dates. That is, BRF is the best fit for all 366 degree distributions; for strength distribution, power law may be a better model only for nine dates. We show these nine dates in Table 1 in Methods.

The BRF defines a characteristic scale and divides a distribution in two halves or regimes: below and above the peak. Each regime has a tail that is independently controlled and modeled by a separate parameter: $b$ controls decay on the left side where nodes with small strength or degree fall, and $a$ regulates decay on the right side where nodes with large strength or degree fall. As the mobility networks evolve with time, so evolves the shape of degree and strength distributions; this is something we can capture with the evolution of fitted $a$ and $b$ parameters. 

\begin{figure}[!t]
    \centering
    \includegraphics[width=1\linewidth]{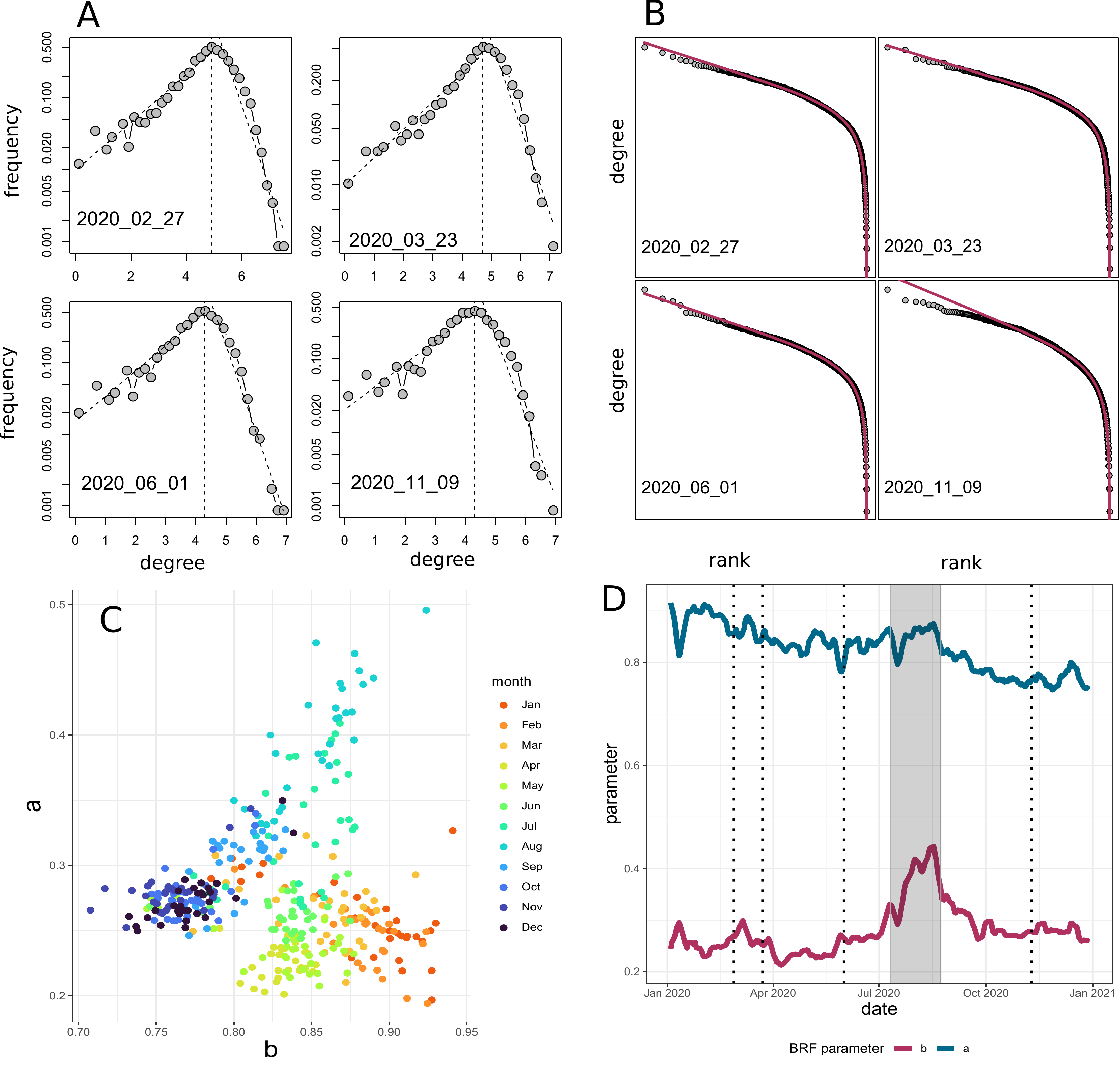}
    \caption{A) histograms (in semilog) of log(degree) for a set of four networks. We also show the mode (peak of the distribution) ans the fitted exponential decays on each side. B) Rank-size plots for degree and fitted BRF for the same networks. C) Scatter plot of fitted $a$ and $b$ parameters of the BRF; each point is coloured according to the month, so we can see the trajectory. D) Time evolution of $a$ and $b$ parameter; we also mark the four relevant dates and the official school vacation period in Mexico. }
    \label{fig::degree}
\end{figure}

We collected data both before and after the beginning of the COVID-19 pandemics, which led to containment measures that altered human mobility. This allows us to analyze the effect of lockdowns on mobility patterns, as derived from degree and strength distribution. For illustration purposes we picked the following set of dates: February 27, first reported COVID case in Mexico; March 23, beginning of official lockdowns; June 1st, beginning of official Epidemiological Stoplight system for the regulation of lockdowns; November 11, beginning of \emph{El Buen Fin}, the largest commercial event in Mexico that took many people out from confinement an into shopping malls and stores.  

\begin{figure}[!t]
    \centering
    \includegraphics[width=1\linewidth]{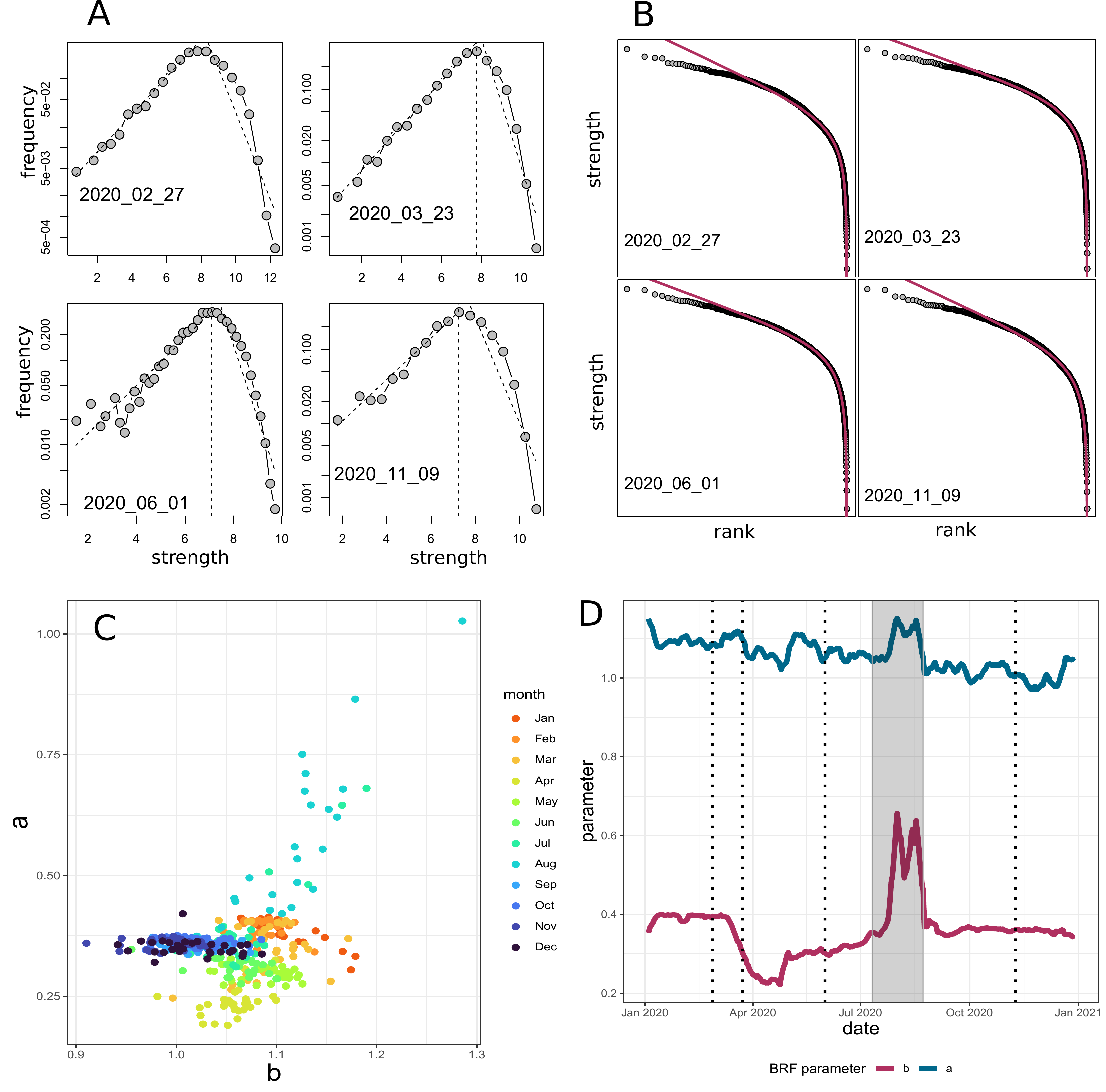}
    \caption{A) histograms (in semilog) of log(strength) for a set of four networks. We also show the mode (peak of the distribution) ans the fitted exponential decays on each side. B) Rank-size plots for strength and fitted BRF for the same networks. C) Scatter plot of fitted $a$ and $b$ parameters of the BRF; each point is coloured according to the month, so we can see the trajectory. D) Time
evolution of $a$ and $b$ parameter; we also mark the four relevant dates and the official school vacation period in Mexico.}
    \label{fig::strength}
\end{figure}

For degree distribution, we show on Fig. \ref{fig::degree} some of these results. First we see on panel A histograms of log(degree) with the y axis in log scale. From these we observe 1) distribution is indeed BRF-shaped and 2) different days have different modes. In panel B we see rank-size plots from these same distributions together with fitted BRF. Panel C shows fitted BRF parameters colored according to month, so we can visually keep track of their evolution. Finally, we show in panel D times series of fitted $a$ and $b$ parameters, which is a different visualization of parameter evolution. We mark the dates we chose for the analysis, as well as official school vacation period, shaded in gray.  Fig. \ref{fig::strength} shows equivalent results but for strength distributions. 

\subsection{Main mobility hubs change with the beginning of lockdowns}

For each node centrality metric (degree and strength) we computed monthly averages and looked at the top nodes for each metric. We call these the mobility \emph{hubs}. Recall that node degree measures the diversity of places from where people move or to where people go from a given node, while node strength is a measure of the total number of people that traveled to or from a given node. 

For degree mobility, we see for example that AGEBs located on and around the main campus of the National Autonomus University of Mexico (\emph{Ciudad Universitaria}) are on the top degree nodes for January, February and March, but they cease to appear on the top from April on. This is the largest university in the Americas in terms of enrollment (more than 360 thousand students, 120 thousand on this campus) and this main campus covers approximately 733 hectares (1811 acres). This is interesting, considering that official lockdowns started on March 23, 2020. We also see that nodes on the Historic Center of Mexico City start appearing among the top starting April. This is the central neighborhood and one of the most important commercial hubs in the city. AGEBs on and around the Mexico City International Airport (AICM) and Chapultepec are consistently on the top, both before and after the beginning of lockdowns. During the study period, the AICM was the only commercial airport in Greater Mexico City and the busiest airport in Latin America in terms of passenger traffic and aircraft movements.  Chapultepec is one of the largest city parks in the world, covering an area of over 686 hectares (1,695 acres). 

For strength mobility, we observe that areas around Santa Fe are among the top for January, February and March, then decay from April to September and reappear on October, November and December. Santa Fe is the largest business district in Mexico City and it holds the largest shopping mall in Latin America, as well as large campus of different private universities. Some areas around hospitals specially dedicated to COVID-19 start appearing among the top during the second half of the year. Finally, areas on and around Chapultepec and the airport consistently appear on the top, before and after the beginning of lockdowns.

\subsection{Nodes with low degree and strength are in the periphery of the city}

\begin{figure}[htp]
    \centering
    \includegraphics[width=1\linewidth]{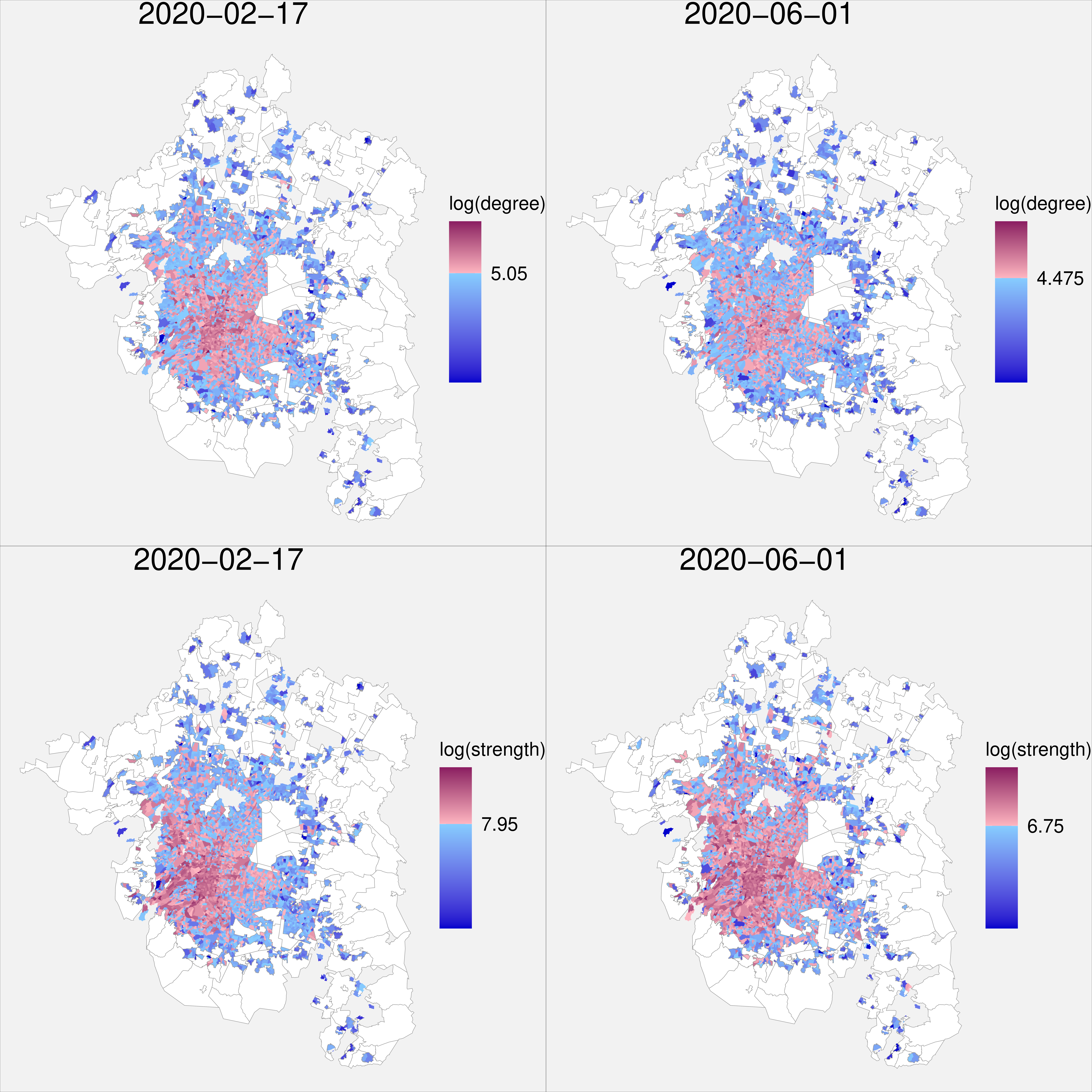}
    \caption{Maps of the AGEBs of Greater Mexico City. Each AGEB is colored according to a given metric (log(degree) or log(strength)). We show in red nodes with a metric above the mode and we show in blue nodes below the mode. White nodes are AGEBs that don't belong to Greater Mexico City. We show by each colorbar the distribution mode for that day.}
    \label{fig::mapa}
\end{figure}

Results so far suggest the existence of a characteristic scale, defined by the distribution mode, and the presence of two regimes in degree and strength distributions: areas below and areas above the mode. A natural question arises regarding geographic location of nodes in these two regimes. We show in Fig.\ref{fig::mapa} maps of Greater Mexico City with AGEBs colored according to degree and strength. Purple correspond to nodes above the mode (high mobility areas), while blue correspond to nodes below the mode (low mobility areas). Regions in white are neighboring areas that either are not urban AGEBs or they don't officially belong to Greater Mexico City.  For comparison purposes,  we chose a date before and a date after the beginning of lockdowns (February 17 and June 1st, respectively). Both of these dates are Mondays, so we can make a more or less fair comparison.  Here we can also see how the value of the distribution mode changes from day to day.  What we can almost immediately see is how the majority of high mobility areas are in the central regions of the city, while low mobility areas tend to be distributed in the periphery. 

\subsection{High/low degree nodes also have high/low strength; exceptions occur in the periphery}

\begin{figure}[htp]
    \centering
    \includegraphics[width=1\linewidth]{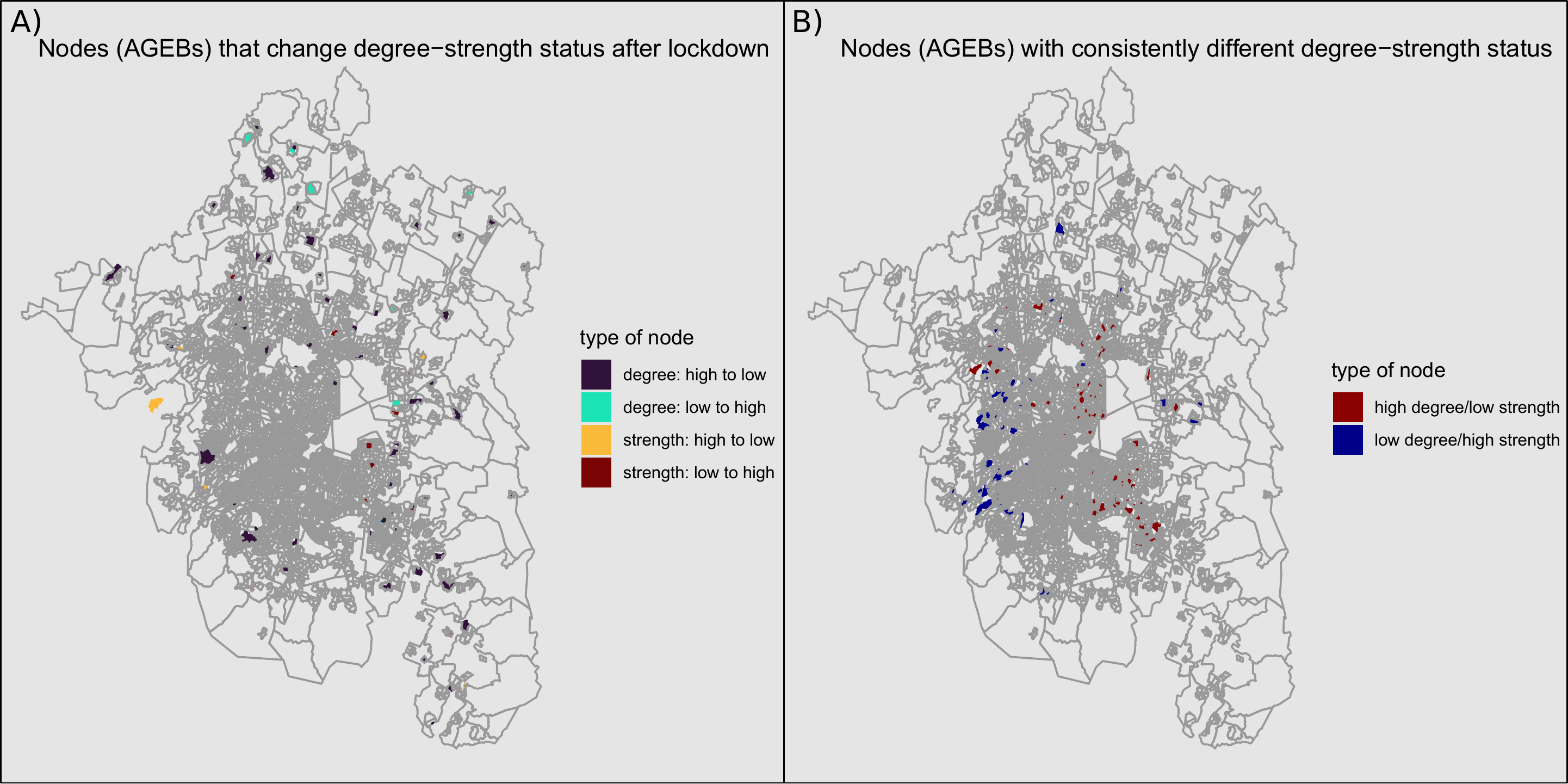}
    \caption{A) AGEBs that changed degree or strength centrality group with the beginning of lockdowns. B) AGEBs that for more than 200 days are on different degree and strength groups on the same day.}
    \label{fig::groups}
\end{figure}

There are 5705 urban AGEBs on Greater Mexico City, so this is the maximum number of nodes that our origin-destination networks can have (network size slightly fluctuates from day to day, because some days there are some areas where we have no observations).   Each node (AGEB) during each day belongs to a degree group (high or low) and also to a strength group (high or low). We wonder if there are any areas that switched group with the beginning of lockdowns. We take again networks from February 17 (before) and June 1st (after) as references and we show on Fig.\ref{fig::groups}-A the results. As we can see, most of these switching areas are on the periphery of the city. 

Considering all 366 networks, we observe that 88 $\%$ of times nodes belong to the same group for degree and strength. This means that areas with high degree also have high strength and areas with low degree exhibit low strength 88 $\%$ of the time. We observe a total of 155 nodes that consistently (for more than 200 days) have different degree strength statuses on the same day, either high degree and low strength or low degree and high strength. We show these areas on Fig.\ref{fig::groups} B). Again, the majority of these areas are on the periphery. 

\subsection{Highly populated areas with low marginalization level show higher mobility metrics}

Results on Fig.\ref{fig::mapa} suggest that proximity to the city center has an impact on degree and strength. We wonder if other kind of variables have such an impact. Additionally to distance to the center (geographic component), we also analyzed total population (demographic component) and social marginalization index (socioeconomic component). Total population data were taken from the 2020 National Census (\url{https://www.inegi.org.mx/programas/ccpv/2020/}). Social Marginalization Index is an official metric, developed by the National Population Counsil (CONAPO) attempting to differentiate urban AGEBs according to shortcomings in education, health, dwelling and  services (\url{https://www.gob.mx/conapo/documentos/indices-de-marginacion-2020-284372}). There are five levels of social marginalization: very high, high, medium, low and very low. Finally, distance to city center was measured as the euclidean distance of each AGEB centroid on the shape file to the main city square, known as \emph{el Zócalo}.  

\begin{figure}[htp]
    \centering
    \includegraphics[width=1\linewidth]{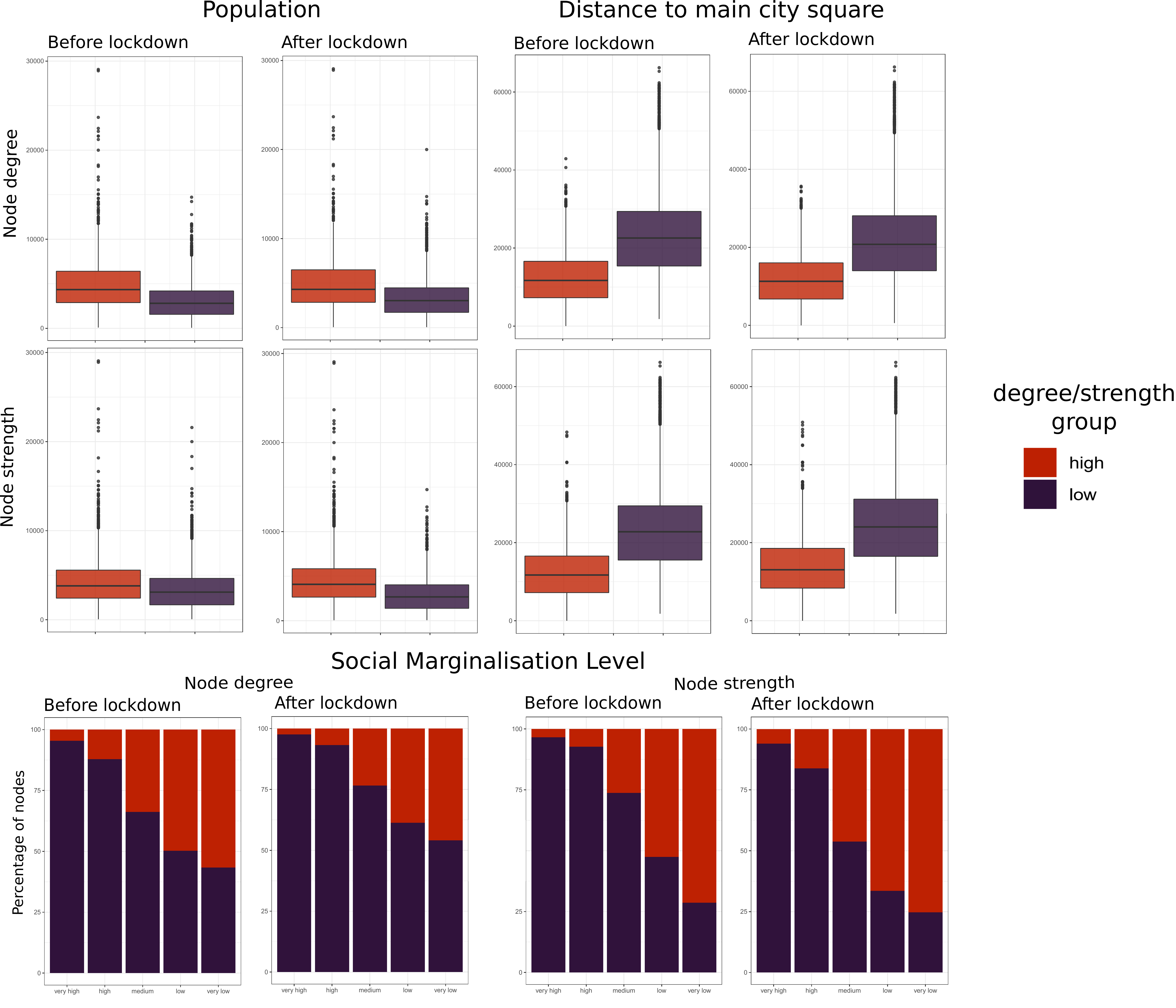}        
    \caption{Top: AGEB population and distance to city main square box plots for high and low degree and strength groups, both before and after the lockdowns. Bottom: for each social marginalization level, percentage of AGEBs that are in the high and low degree and mobility groups. }
    \label{fig::population}
\end{figure}

We show the main results of this analysis on Fig. \ref{fig::population}. For each metric (degree and strength) we partition the nodes in high (red) and low (blue) level and we show a box plot of total population and a box plot for distance to main city square. In each case, we utilize the reference networks before (February 17) and after lockdown (June 1st). For each of these eight cases we performed Wilcoxon signed rank tests and confirmed in all cases (with p-val $< 2.2e^{-16}$) that high population distributions are shifted to the right and short distance distributions are shifted to the left. For the case of social marginalization, we show for each level the percentage of nodes that are in the high and low mobility groups.  Additionally, we fitted logistic regression models (\emph{mobility group} $\sim$ \emph{population}, \emph{mobility group} $\sim$ \emph{distance} and \emph{mobility group} $\sim$ \emph{social marginalization level}), both before and after the beginning of lockdowns. In all cases we found statistical significance of the regressor variable with p-val $< 2.2e^{-16}$. 

\section{Discussion}

The first thing we notice in this set of networks is that in- and out-degree are almost always strongly correlated. The same thing happens with in- and out-strength, except for a few number of anomalous nodes that tend to be geographically together and on the periphery of the city. Therefore, we focused our investigation on total-degree and total-strength. A deeper analysis of the subtle but interesting differences between in- and out is a matter of future study. 

One of the first findings of this investigation is that the origin-destination networks for Mexico City are not scale free. Usually, a network is said to be scale free if the degree distribution follows a power law; since a power law is invariant under rescaling, it is said that the network is ``free'' of a natural scale \cite{broido2019scale}. It depends on the phenomenon being modeled, but the usual interpretation of this is that there is no natural scale for the phenomenon and that elements of all sizes are governed by the same rules. This is not the case for our set of origin-destination networks, where neither degree nor strength are well fitted by a power law. Instead, we see that degree and strength have unimodal distributions with two tails, each one approximately obeying a different power law. This means that there are at least two characteristics scales on the networks: nodes with high centrality (degree or strength) behave differently than nodes with low centrality. This is what motivates us to propose the existence of two different mobility groups or regimes, high and low, depending on whether the centrality metric is above or below the mode. This is something that we can model with the BRF distribution and it allows us to ask different questions about human mobility in the city. 

For example: are degree mobility and strength mobility the same thing? Recall that degree centrality measures the number of places connected to a given node, while strength centrality attempts to measure the number of people that travel to or from a given node. Our results show that high degree implies high strength (and low degree implies low strength), but not always. Exceptions occur mainly on the periphery of the city, as shown in Fig.\ref{fig::groups}-B. Interestingly, low degree/high strength areas occur on the west part of the city, while high degree/ low strength occur on the east. This corresponds to a growing urban dynamics in Mexico City where the east has experienced uncontrolled urban growth, with a mixture of industrial and residential land use, high population density and very low income levels, while the majority of residential high-income neighborhoods are on the west part of the city (see for example \cite{graizbord2007movilidad,moreno2011factores}). In a way, people on the west move to many different places in the city, while people on the west move a lot inside their respecting and surrounding neighborhoods. 

These observations motivate other questions, such as: do social, economic and geographic factors correlate with mobility? In terms of our high/low mobility division, we can restate the question as: do these factors statistically explain being on one mobility group or the other? The answer is: yes, they do. Population distribution of high and low mobility groups, both for degree and strength, are significantly shifted, that is high mobility areas have higher populations than low mobility areas. The same happens with distance to main city square: areas with high mobility tend to be closer to the city center. In a similar way, areas with high mobility metrics have lower levels of social marginalization. Furthermore, these shifts and differences are not evidently different before and after the lockdowns, suggesting that lockdowns affected both mobility regimes more or less equally. 

With the aid of the BRF distribution, we can see in more detail how each mobility regime changes during the study period. Recall that $b$ parameter controls the left tail (low mobility) and $a$ the right tail (high mobility). Now look at time series of $a$ and $b$ in Figs.\ref{fig::degree}-D and \ref{fig::strength}-D. For degree mobility, both time series have little correlation: $b$ has a slight and sustained downward trend, while $a$ has a peak during the summer vacation period, followed by a more or less flat behavior. An increase on each parameter translates in a decrease in the slope of its corresponding tail, that is, a slower decay. We see, for example, that during summer vacation the weight of the right tail increases, indicating a larger variance and a greater presence of outliers (nodes with very high degree). These changes are more evident for the strength distributions. Here both parameters suffer a decay that coincides with the beginning of lockdowns (decay in $a$ is quite abrupt) and a very high peak during summer vacation. The interpretation that we give is the following: with the lockdown, mobility was greatly reduced and strength distribution became narrower  - extreme nodes are now rare - but with summer vacation a wide variety of values for node strength reappeared, with extreme values (but high and low) being common again. Interestingly, summer vacation seems to affect these mobility patterns more than the initial lockdown.

On a more descriptive way, we observe that top mobility hubs changed with the start of the pandemic. There are some areas, such as the airport and the city park of Chapultepec, that were persistently among the top ranked areas in terms of our mobility metrics. We identify these two places as the main mobility hubs in Mexico City during 2020. In the case of the airport, there was a drop on the total number of incoming flights, from approximately 13 thousand during March to approximately 2 thousand in April (\url{https://www.datatur.sectur.gob.mx/SitePages/TrasnAerea.aspx}). Pre-pandemic levels were not recovered during 2020. In spite of this, the airport remains as a mobility hub (here we mention that Mexico had a very loose border restriction, with no need for international travelers to provide a negative COVID test or proof of vaccination \cite{leyva2022cierre}). With many people reducing their mobility during the first months of the pandemic, many of the people that keep moving is people that go to the airport. In a similar way we can explain why areas on the historic center of Mexico City rise to the top ranked nodes with the beginning of the lockdown. Being one of the main commercial and services areas in the city, we see that people who keep moving in the city during the lockdowns are people that go to this place, presumably people that work on provision of products. To look into this with more detail is a matter of future investigation. All together, we identify the historic center (commercial district), Chapultepec (city park), Santa Fe (business district) and the airport as the main mobility hubs on the city.

Finally, the BRF function provides a simple way to track the evolution of a distribution, by looking at the values of $a$ and $b$. We see on scatter plots of Figs.\ref{fig::degree} and \ref{fig::strength} that did not return to pre-pandemic patterns during the study period (we are talking about centrality distribution, not a quantity of total mobility).

\section{Conclusions and perspectives}

Mexico City is a very large metropolitan area where urban mobility patterns are heterogeneous and complex. There are many things we can learn about these mobility patterns by looking at node degree and strength on origin-destination networks. These two centrality metrics aim to measure two distinct aspects of mobility: node degree measures the variety of places from where people arrive to the node or to where people go from the node, strength centrality measures the total number of travels to or from the node.

Origin-destination networks for Mexico City during 2020 are not scale free, centrality distribution peak defines a characteristic scale and they exhibit two regimes of behavior (both for degree and strength) divided by the mode. While nodes with high/low degree tend to have high/low strength, this is not always the case. Exceptions, as well as nodes that switch group with the beginning of the pandemic, mainly occur on the periphery. Some of the patterns that we observe on this regard correspond to a growing and well studied narrative in Mexico City of ``rich west / poor east''. 

High mobility areas tend to have higher population, lower marginalization levels and tend to be closer to the center of the city. The outburst of COVID-19 and its lockdowns in 2020 altered mobility patterns in the city; high and low mobility regimes were not always altered in the same way. 

The set of origin-destination networks that we analyzed was constructed from a large data base of anonymized cellphone geolocation data and allow for the formulation of many other interesting questions: what is the community structure of these networks? Can we use this community structures to delineate functional regions in the city? How do network metrics relate with new cases and deaths of coronavirus? Perspectives of this work including providing answers to all these questions. 

\section{Methods}
\label{appendix}

\subsection{Network analysis}
Specifics for the methodology for geolocation data collection, curation and network construction is presented in \cite{arXiv}. The result is the set of origin-destination networks provided in the Open Science Framework (OSF) repository cited below. Networks are provided in csv files in edgelist format. These directed networks were analyzed  with the \emph{igraph} library for the R programming language \cite{csardi2013package}.    

\subsection{Model fitting and goodness of fit tests}

\begin{figure}[htp]
    \centering
    \includegraphics[width=1\linewidth]{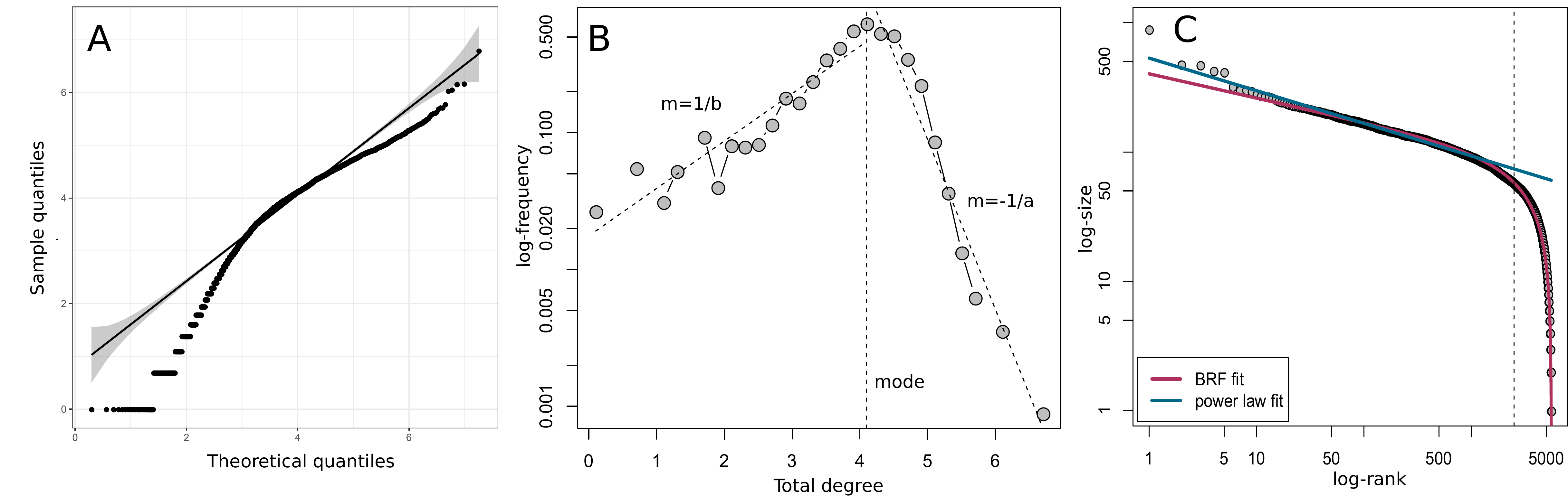}
    \caption{Illustration of our methodology for model fitting and selection. A) Q-Q plot for log(degree), B) histogram of log(degree) (in semilog) and C) rank-size function of degree. These degree distribution correspond to the network of April 26.}
    \label{methodology}
\end{figure}

In a weighted network the degree of a node is the total number of adjacent edges (its total number of neighbors) and the strength is the sum of weights of adjacent edges. In directed networks, where we distinguish between in and out degree and strength, total degree/strength of a node is the sum of in and out degrees/strengths. For the total-degree distribution and the total-strength distribution and for each of the 366 mobility networks we compare the performance of three statistical models: lognormal, power law and BRF. We show an example of our procedure for the total-strength distribution of the April 26 network. First we perform a q-q plot for the logarithm of total degree (Fig.\ref{methodology}-A). Here we see the lognormal distribution is discarded, as there are significant deviations from normality at both tails. This is confirmed with a Kolmogorov-Smirnov (KS) test (p-val $< 2.2e-16$). Then we find best fits for power law and BRF by a non-linear regression (Levenberg-Marquardt algorithm). We perform KS tests also for the power law and the BRF fits. In this example, values for the KS statistic for lognormal, power law and BRF are 0.984, 0.247 and 0.0512 respectively. We see that, according to this test, BRF is the better model. However, BRF is a generalization of power law,  so a better fit is expected just as a consequence of the additional parameter. We compare the Akaike Information Criterion (AIC) on both models, which is an estimator of prediction error that considers the different number of parameters. For this example, the difference in AIC for both models is $\Delta AIC = AIC_{BRF}-AIC_{power law}=-11006$, indicating that BRF yields a better fit (less information is lost). We show on Fig.\ref{methodology}-B the histogram of log(total-degree) in semilog representation and can observe the behavior of the log-BRF density, that is, a smooth peak with two tails that decay independently of each other, the lower tail controlled by $b$ and the upper by $a$. Finally, we see on Fig.\ref{methodology}-C the rank-size plot of the observations, together with the best BRF fit and the best power law fit. These visualizations, together with KS tests and AIC, indicate that BRF is the better of the three proposed models. We chose April 26 to illustrate our methodology for clarity sake, because it is one of the best fits to BRF. \\

\begin{table}[t]
\label{table:fits}
\centering
\begin{tabular}{lrrrr}
  \hline
  date & KS (power law) & KS (BRF) & KS (lognormal) & $\Delta$Akaike \\ 
  \hline
 2020-01-14 & 0.17 & 0.19 & 1.00 & -10707.66 \\ 
   2020-01-28 & 0.18 & 0.18 & 1.00 & -10891.57 \\ 
   2020-02-25 & 0.18 & 0.19 & 1.00 & -11020.94 \\ 
   2020-02-28 & 0.17 & 0.17 & 1.00 & -11124.51 \\ 
   2020-03-04 & 0.18 & 0.19 & 1.00 & -10793.35 \\ 
   2020-08-02 & 0.21 & 0.21 & 1.00 & -11150.23 \\ 
   2020-08-11 & 0.19 & 0.31 & 1.00 & -11490.34 \\ 
   2020-08-16 & 0.20 & 0.25 & 1.00 & -12549.98 \\ 
   2020-08-20 & 0.19 & 0.59 & 1.00 & -11856.50 \\ 
   \hline
\end{tabular}
\caption{For nine particular dates, power law is a better fit than BRF for strength distribution according to the KS test. For these same dates, BRF has a better AIC than power law.}
\end{table}

\vspace{1cm}
\break

\textbf{Acknowledgments.}
This project was supported by the Fondo Conjunto de Cooperación México-Uruguay from the Agencia Uruguaya de Cooperación Internacional and  the Agencia Mexicana de Cooperación Internacional para el Desarrollo.

\textbf{Competing interests.} The author have no competing interests to declare.

\textbf{Data availability.} The 366 origin-destination networks dataset is deposited in the Open Science Framework (OSF) repository in edgelist format (DOI 10.17605/OSF.IO/639ZX). Data regarding AGEBs population and social marginalization levels are publicly available on the urls cited in the manuscript. This will enable reader to replicate all the results presented in this paper.


\bibliographystyle{unsrt}
\bibliography{references}

\end{document}